\documentclass[11pt,twoside]{article}
\usepackage{asp2010}

\resetcounters

\begin{document}

\title{White Light Flare Continuum Observations with ULTRACAM}
\author{Adam F. Kowalski$^1$, Mihalis Mathioudakis$^2$, Suzanne L. Hawley$^1$, Eric J. Hilton$^1$, Vik S. Dhillon$^3$, Tom R. Marsh$^4$, Chris M. Copperwheat$^4$
\affil{$^1$Astronomy Department, University of Washington Box 351580,
  Seattle, WA 98195, USA; adamfk@u.washington.edu}
\affil{$^2$Astrophysics Research Centre, School of Mathematics and Physics, Queen's University, Belfast BT7 1NN, UK}
\affil{$^3$Department of Physics, University of Sheffield, Sheffield S3 7RH, UK}
\affil{$^4$Department of Physics, University of Warwick, Gibbet Hill Rd, Coventry CV4 7AL, UK}}

\begin{abstract}
  We present sub-second, continuous-coverage photometry of three flares on the
  dM3.5e star, EQ Peg A, using custom continuum filters with
  WHT/ULTRACAM.  These data provide a new view of flare continuum
  emission, with each flare exhibiting a very distinct light curve morphology.
  The spectral shape of flare emission for the two large-amplitude
  flares is compared with synthetic ULTRACAM measurements taken from the spectra during the
  large `megaflare' event on a similar type flare star.  The white light
  shape during the impulsive phase of the EQ Peg flares is consistent
  with the range of colors derived from the megaflare continuum, which is known to 
  contain a Hydrogen recombination component and compact,
  blackbody-like components.  Tentative evidence in the ULTRACAM
  photometry is found for an anti-correlation between the emission of these components.
\end{abstract}

\section{Introduction}
Active M dwarfs are notorious for white light flares in the near-UV
and optical. 
White light emission is one of the first types of radiation to appear during both solar 
and stellar flares \citep{NeidigKane1993, Hawley1995} with 
 a quickly-evolving impulsive phase followed by an extended
 decay phase that can sometimes last up to several hours
 \citep{HawleyPettersen1991, Kowalski2010a}.  Time-resolved spectral observations of M
 dwarf flares show a rising continuum into the near-UV, which has been
 modelled as a hot,  $T \sim$ 10,000 K blackbody spectrum
 \citep{HawleyFisher1992, Fuhrmeister2008}.  Recent  spectral observations
 have revealed that a Hydrogen recombination component (Balmer
 continuum) can also contribute significantly to the decay phase of the 
white light for $\lambda < 3646$\AA\ \citep{Kowalski2010a}.

Spectral observations can be severely limited by time resolution, with integration times (minutes) typically
much longer than the timescales of flare evolution (seconds);  as a result, broadband
filters are more often used to characterize the time-evolution of the
white light continuum.  An analysis of 4 flares on the dM3e star AD Leo
using UV$+UBVR$ filters has shown a nearly isothermal $T \sim
8500-10,000$ K component in the white light at all times
\citep{Hawley2003}.  
Another study by \citet{Zhilyaev2007} analyzed high time-cadence ($\sim$0.1s)
$UBVRI$ photometry and concluded that multiple components
(a Hydrogen recombination component and $\sim$18,000 K blackbody)
contribute to varying degrees during a flare on the dM3.5e star EV Lac.

The origin of the white light continuum during flares is a long-outstanding problem 
in solar and stellar
flare physics.  The \citet{Allred2006} non-LTE radiative hydrodynamic
(RHD) models of M dwarf flares predict two continuum sources, but the
photosphere does not receive enough energy to generate the temperatures 
near $\sim$10,000 K implied by the observations.
However, the model white light emission resembles the shape of a hot ($T\sim8900$ K) blackbody when 
convolved with the UV$+UBVR$ filters \citep[see Figure 12 of][]{Allred2006}, indicating that there might be a degeneracy in the measured
flare shape when using broad filters that encompass numerous spectral
lines and continua.

We have begun a flare monitoring campaign with ULTRACAM using custom
continuum filters to
characterize the white light continuum properties on sub-second
timescales.  In these proceedings, 
we present our first ULTRACAM observations of the flare star EQ Peg A.

\section{Observations}
 
ULTRACAM is an ultrafast, triple-beam, cascade dichroic CCD camera,
and has helped open up simultaneous multiband, sub-second time domain
astronomy \citep{Dhillon2007}.  On UT 10 August 2008, we obtained flare observations of EQ Peg
A and B using ULTRACAM on the 4.2-m William Herschel Telescope at La
Palma.  EQ Peg A and B are a resolved binary system consisting of
dM3.5e and dM4.5e components, which have well-studied flare
properties from the X-ray through the optical \citep{Lacy1976, Robrade2004,
  Mathioudakis2006, Liefke2008}.

The observations were taken with an H$\alpha$ filter (FWHM $\sim $ 50\AA) in addition to two custom 
continuum filters, NBF3510 ($\lambda_c \sim$ 3510\AA, FWHM $\sim
100$\AA) and NBF4170 ($\lambda_c \sim$ 4170\AA, FWHM $\sim
50$\AA), designed specifically to avoid line emission and the Balmer
jump at $\lambda=3646$\AA.  Figure 1 (left panel) shows  a flare
spectrum from \citet{Kowalski2010a} with the NBF3510, NBF4170, and
Johnson $U$ band transmission efficiencies.  The $U$ band contains 
high-order Hydrogen emission and straddles the Balmer jump 
whereas the custom filters effectively isolate the continuum emission.
Exposure times were 0.291 s for the NBF3510 camera and 0.134 s for the
NBF4170 and H$\alpha$ cameras.

The data were reduced with the ULTRACAM pipeline
software\footnote{http://deneb.astro.warwick.ac.uk/phsaap/software/ultracam/html},
and we used a nearby comparison star, Gl 896 C, to obtain differential photometric measurements in all bands.  A fixed aperture radius of 6 pixels was used.  The night was photometric with variable seeing, and the observations were terminated due to morning twilight.  

\section{Light Curve Data}
During the last $\sim$30 minutes of observations, EQ Peg A produced
three flares, each with a unique light curve morphology (Figure 1,
right panel):  a
`gradual' flare (\textbf{G}) at $t \sim 50$ min, an `impulsive' flare
(\textbf{I}) at $t \sim 64$ min, and a `traditional' flare
(\textbf{T}) at $t \sim 70$ min.  The high-cadence, continuous
coverage photometry allows us to analyze the time-evolution of these
flares without the ambiguity imposed by integration and readout times,
which can be much greater than the timescales of flare
evolution.  We speculate that the differences in these light curve
morphologies 
could be due to different active region structures
(i.e., small vs. large arcades of flaring loops), different
non-thermal particle beam flux evolution, and/or different white light continuum properties.

In general, each flare has similar morphologies in the
NBF3510 and NBF4170 filters, and Table 1 summarizes the key
properties.  
The \textbf{G} flare contains several emission peaks and
a decay time that is long for its small peak amplitude.  
The \textbf{T} flare has a typical fast-rise,
exponential-decay (`FRED') shape, but exhibits several variations from this
trend.  The \textbf{I} flare is nearly symmetric about the peak, but a low-amplitude, short
decay is apparent when examined over a small time window. It is
interesting that the \textbf{T} and
\textbf{I} flares have similar peak amplitudes, but the \textbf{T}
flare has nearly 5x the equivalent duration (energy) as the \textbf{I} flare.

\begin{figure}[!ht]
  \plottwo{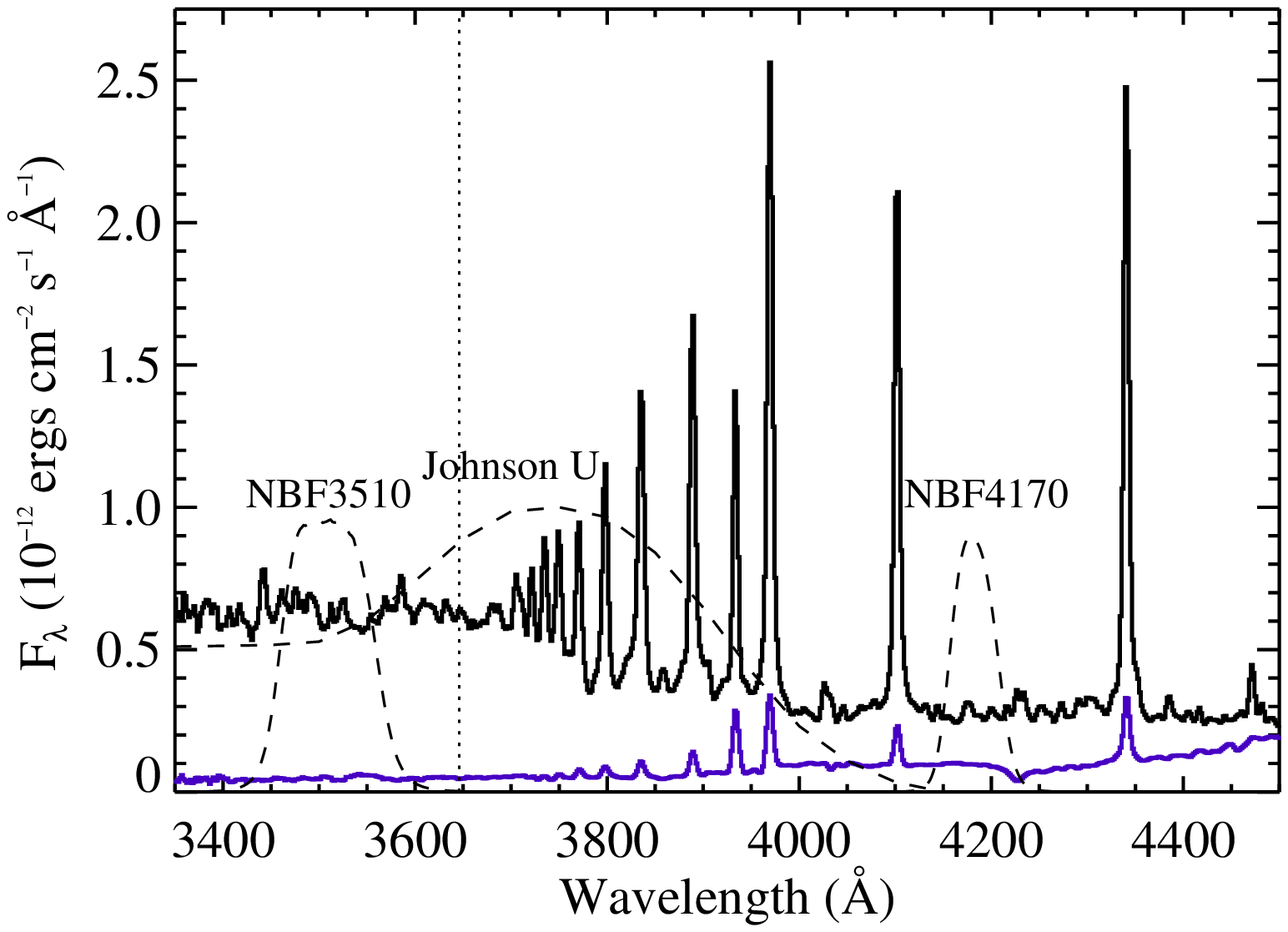}{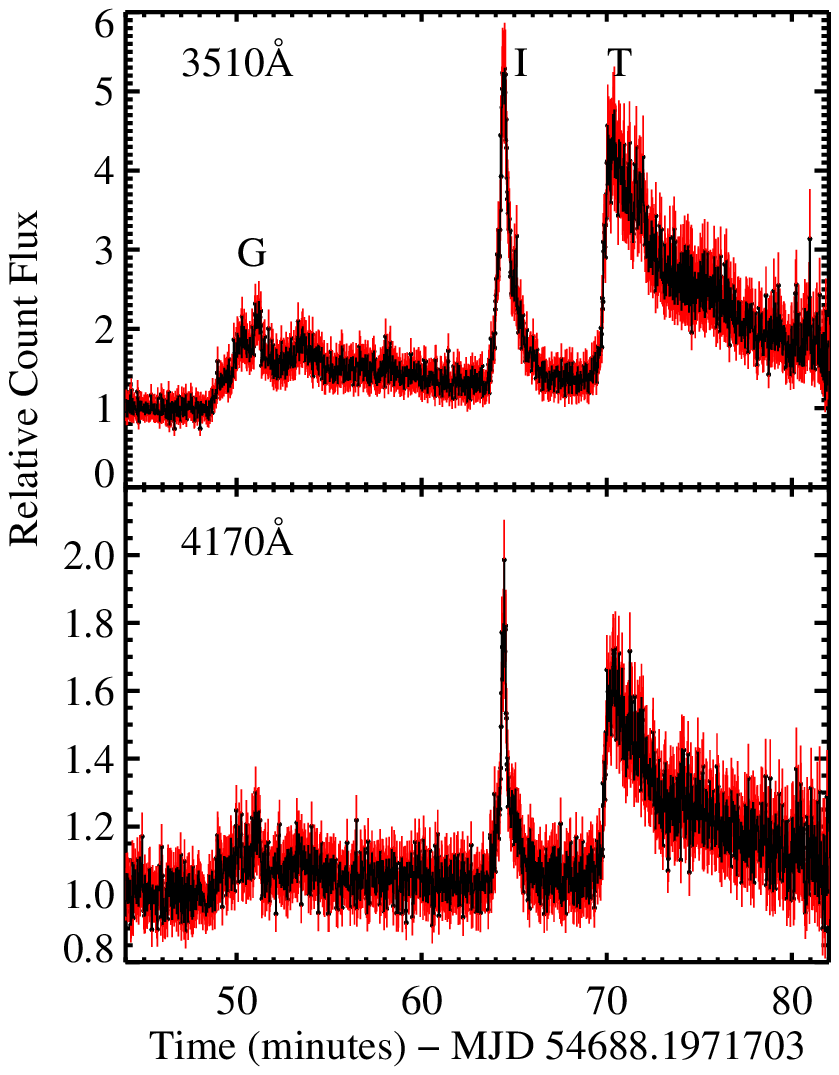}
\caption{Left -- Example flare (black) and quiescent (purple;
 scaled by a factor of 3) spectra from
  \citet{Kowalski2010a} with the transmission curves of the NBF3510,
  NBF4170, and the \citet{MaizA2006} Johnson U band.  The Balmer
  jump wavelength is indicated by a vertical dotted line.  
 Note that the flare-to-quiescent contrast is much greater in the
 NBF3510 filter.  Right -- Three flares on EQ Peg A in the NBF3510 and
 NBF4170 filters.  We show
every 4th point (top) and every 8th point (bottom) for clarity.}
\end{figure}

\begin{table}[!ht]
\caption{NBF3510 Flare Properties}
\smallskip
\begin{center}
{\small
\begin{tabular}{cccc}
\tableline
\noalign{\smallskip}
Flare type & peak amplitude ($\sigma$) & Decay Time Constant [min] & Equivalent
Duration [min] \\
\noalign{\smallskip}
\tableline
\noalign{\smallskip}
gradual (\textbf{G}) & 2.1 (0.1) & 14 & 7.6 \\
impulsive (\textbf{I}) & 5.1 (0.3)  & 1 & 3.2 \\
traditional (\textbf{T}) & 4.3 (0.3)  & 7 & 14.8 \\
\noalign{\smallskip}
\tableline
\end{tabular}
}
\end{center}
\end{table}

\section{Flare Color Analysis}
If the underlying spectral energy distribution of the
star is known, then the relative count light curves of Figure 1 (right
panel) can be 
converted to a light curve of the spectral shape of the flare
emission.  Flux-calibrated M dwarf spectra near the atmospheric cutoff
are rare, so we obtained deep 3400 - 5500\AA\ spectra of EQ Peg A \& B
with the ARC 3.5-m at the Apache Point Observatory in October,
2008. The spectra were reduced and flux-calibrated using standard
IRAF\footnote{IRAF is distributed by the National Optical Astronomy Observatories,
which are operated by the Association of Universities for Research
in Astronomy, Inc., under cooperative agreement with the National
Science Foundation} procedures. Following the spectrophotometry
procedures of \citet{Sirianni2005} and \citet{MaizA2006}, we
estimate the ratio of quiescent fluxes of EQ Peg A in the NBF3510 and
NBF4170 filters to be 0.40, with $\sim$5\% uncertainty.  If $C(t)$ is
the relative count flux (normalized to 1 during quiescence), $C_o$ is the relative count flux
immediately before the flare, and $R_Q$ is the quiescent spectral ratio ($
F_{3510,Q} / F_{4170,Q}$), then the spectral shape of flare emission
around the Balmer jump is 
\begin{equation}
R_Q \times \frac{ C_{3510}(t) -C_{o,3510}}{C_{4170}(t)-C_{o,4170}}.
\end{equation}
 This formula follows from the integrand
of the equivalent duration \citep{Gershberg1972} for two filters, 
and we refer to it as the \textit{flare color}.

\subsection{The Flare Color as a Proxy for Two Continuum Components}
Recent spectral observations of a `megaflare' on the dM4.5e
star, YZ CMi, revealed an anti-correlated time-evolution between the
Balmer continuum (BaC) and a $T \sim 10,000$ K blackbody
continuum components \citep{Kowalski2010a}, which 
can be explained by a Balmer jump in absorption forming 
during the secondary flares \citep{Kowalski2010b}. 
Figure 2 shows how this flare would appear
in the ULTRACAM continuum filters, using synthetic photometry
and flare colors calculated from the spectra.  The NBF3510 emission
contains both blackbody and BaC components, whereas the NBF4170
contains primarily blackbody emission (and also probably a small
amount of Paschen continuum; see \citet{Allred2006}).  The anti-correlation 
between the total flare emission (U band,
synthetic NBF3510, synthetic NBF4170) and the flare color is prominent throughout the
entire decay and reflects the changing contributions of the blackbody and
BaC components.  

\begin{figure}[!ht]
\plotone{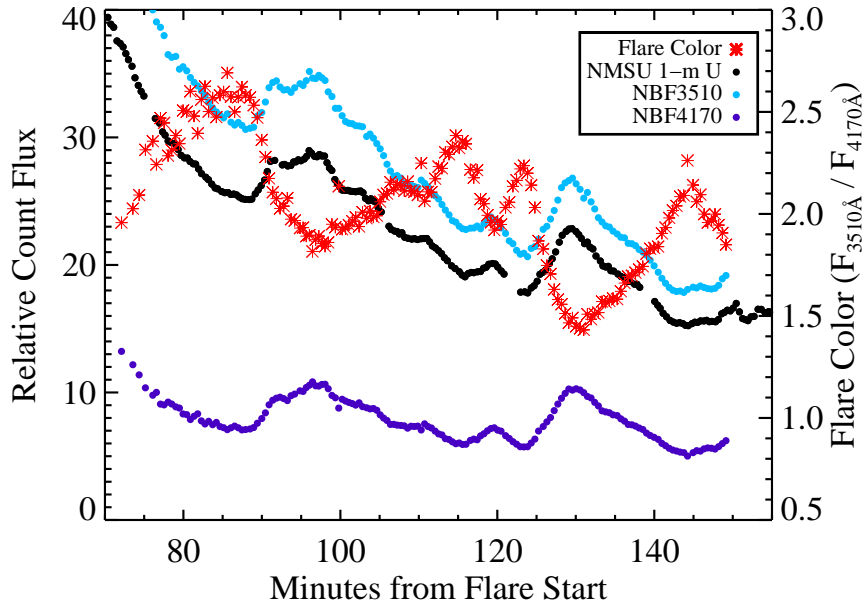} 
\caption{The U band light curve for 1.3 hrs of the
  megaflare on the dM4.5e star YZ CMi (see \citet{Kowalski2010a} for
  the complete light curve).  From the simultaneous optical ARC 3.5-m
  spectra, we calculated synthetic NBF3510 and NBF4170 light curves and
  the flare color.  An anti-correlation 
is prominent between the relative count fluxes and the color
evolution.  Note also the overall decline in flare color.}
\end{figure}

\begin{figure}[!ht]
  \plotone{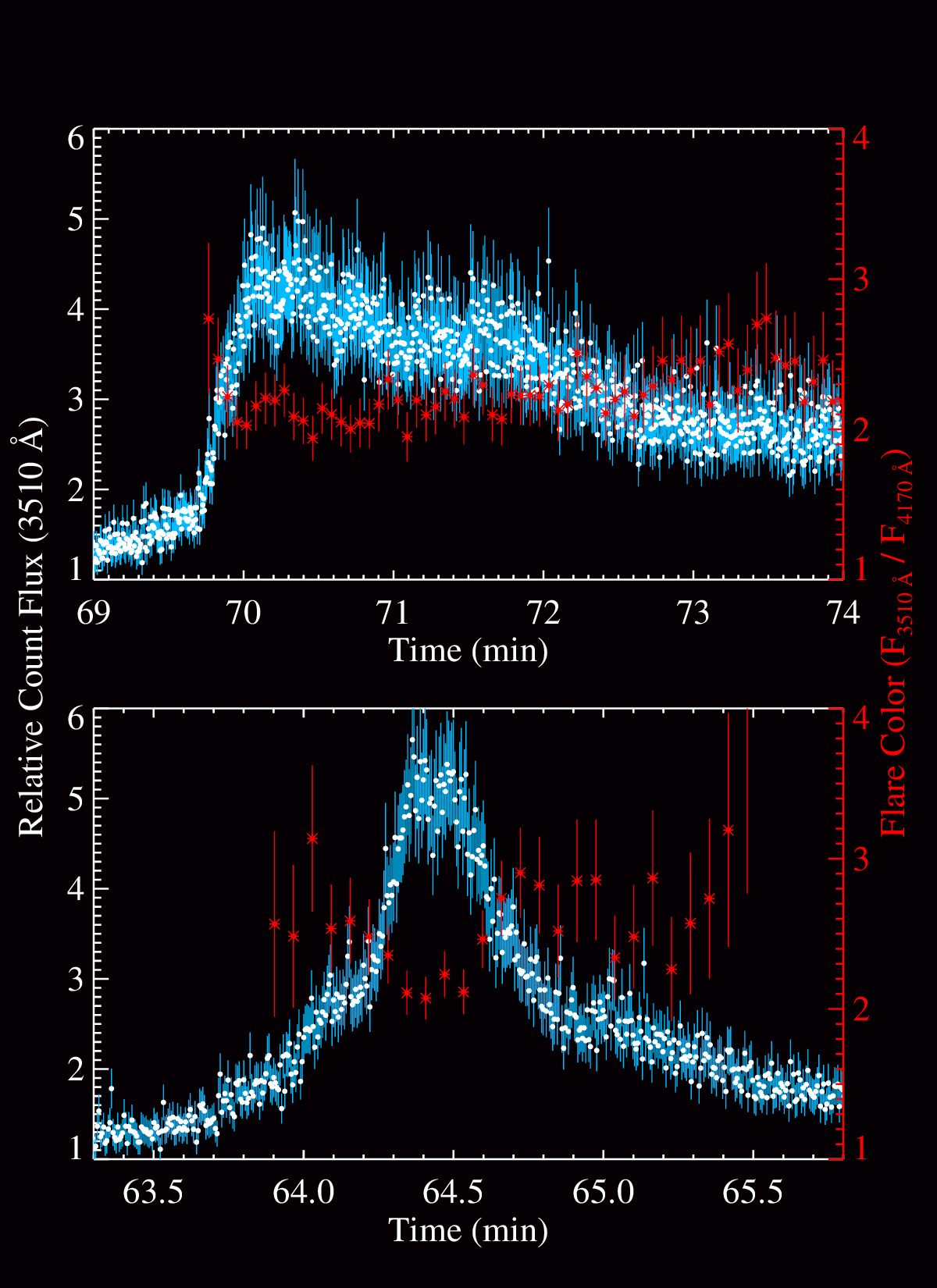}
\caption{A closer view of the NBF3510 data for the \textbf{T} flare (top
  panel) and \textbf{I} flare (bottom panel) reveals a variety of substructure 
beyond a simple fast-rise, exponential decay shape. At this
time-resolution, the rise in both flares
has several distinct phases.  The flare color is shown in red (right axis) and is binned by
3.8 s to reduce the errors.  Note that the \textbf{T} flare has a decay phase
that extends past the end of the observations at $t \sim 83$ min.  }
\end{figure}

For fast time cadences not achievable with spectral measurements, the
ULTRACAM flare color acts as a proxy for the relative 
contributions of the two white light continuum components.    
The flare color evolution is shown for
the \textbf{I} and \textbf{T} flares in Figure 3.  The errors are obtained by
propagating the pipeline uncertainties, the standard deviation of the
relative count flux in the preflare level, and the uncertainty in the
quiescent spectral shape.  The low-level flux enhancements during most
of the flaring times lead to large errors for the original cadence.
Binning the flare color into $\Delta t = 3.8$ s bins allows for a time-resolved 
analysis during the flare peak times.  

The flare colors of the \textbf{I} and \textbf{T} flares have several
notable similarities to the megaflare.  The average flare color
around the main impulsive phase of the \textbf{I} and \textbf{T} flares is $\sim$2.2, which falls within the range of flare colors
derived from the megaflare spectra, $\sim$1.5 - 2.5.  The \textbf{I} and
\textbf{T} flares have much smaller amplitudes than the megaflare, which
was emitting at very high levels, $\sim$40 times the
quiescent level, during the spectral observations.  However, the
similar flare colors imply 
a common geometrical scaling of the continuum components near the peaks of the
\textbf{I} and \textbf{T} flares and during the decay phase of the megaflare, with the
blackbody source having
$\sim$1/10 the area of the BaC-emitting region.

We find tentative evidence 
for an anti-correlation between the flare color and the 
NBF3510 light curve at several times during the $\textbf{I}$ and
$\textbf{T}$ flares.  
This is especially evident during the fast rise and fast decline phases of the \textbf{I}
flare, which exhibit similar anti-correlated variations as in the
secondary flares at $t \sim95$ min and $\sim130$ min in Figure 2.  
If all continuum components varied by the same factor (i.e.,
if the increasing flare flux resulted from scaled up 
emitting area of all components), then the flare color would be flat
as a function of time.  The changing flare color likely indicates the varying contributions of each
continuum component, such that the blackbody has its largest
relative contribution during the peak phase.  Whether or not the
decrease in flare color also results from the apparent amount of BaC emission
decreasing as in the megaflare -- due to increased Balmer continuum in 
\emph{absorption} from the blackbody-like component
\citep{Kowalski2010b} -- is difficult to determine from two-band photometry. Nonetheless, 
the observed flare color of $\sim$2.2 is consistent with this scenario. 

Despite the similarities between the \textbf{I} and \textbf{T} flares
and the megaflare, one can see that the flare color in Figure 2
exhibits an overall decrease, whereas a decreasing flare color is not
found in the decay of the \textbf{I} and \textbf{T} flares.  This
inconsistency is not yet fully understood, but it may be related to the
presence of several secondary (blackbody-like) flares during the megaflare decay.

\subsection{Comparison to Radiative Hydrodynamic Flare Models}
We calculated NBF3510 and NBF4170 fluxes of the F11 flare spectrum from
the \citet{Allred2006} RHD models and found
that the model flare fluxes and colors are inconsistent with the observed and derived properties of 
the \textbf{I} and \textbf{T} flares.  Given an extremely large flare 
that emits from 
5\% of the visible hemisphere (i.e., more than twice the inferred area of
 the BaC-emitting region during 
the megaflare), the model predicts relative peak fluxes that are only 2.0 and 1.07 in the 
NBF3510 and 
NBF4170 filters, respectively.  The discrepancy is particularly
noticeable for the amplitudes of the \textbf{I} and \textbf{T}
flares.  
The \textbf{G} flare has an amplitude that 
is more consistent with the model; however, the required surface area
coverage is likely not realistic for such a short-lived, weak flare.
The flare color of the F11 model spectrum is $\sim$5, 
which is also inconsistent with the observations. 
Increasing the area of the backwarmed photosphere, which might
be a more realistic geometry for a flaring region \citep{Isobe2007}, would
help to decrease the flare color of the model from such a discrepant value.

It is a known problem that the hot blackbody component of the white
light continuum is not produced
in the RHD flare models.   These models
may fall short in matching the observed properties of flares
due to not having enough power in accelerated electrons, and/or 
they may be limited by the one-dimensional treatment of a process that
often includes arcades of flare loops \citep[see
the spatial evolution of solar flare white light footpoints in][]{Wang2007, Wang2009}.  

\section{Conclusions}
We use ULTRACAM continuum observations of EQ Peg A 
to characterize white light flares with custom continuum
filters that do not contain Hydrogen line contamination.  A variety of 
flare light curve morphologies are observed on sub-second timescales, with all of these 
flares exhibiting substructure beyond the standard fast-rise, exponential decay
(FRED) shape. The rate at which each flare shape occurs is not known, but such
 information would help to understand the physical parameters
 and conditions that cause, for example, an \textbf{I}-type flare to occur 
instead of a \textbf{T}-type flare.

We calculate a flare continuum color to quantify the spectral shape
around the Balmer jump.  The flare color evolution shows preliminary
evidence of variations on timescales of several seconds during the impulsive
phase.  The average color over the first peak
of these two EQ Peg flares is consistent with the synthetic flare
colors during the decay phase of a megaflare on the M dwarf YZ CMi, for which we have spectra and therefore a
detailed understanding of each continuum
component.  We find tentative evidence for an 
anti-correlated time-evolution between the relative flux and the flare color, 
similar to that previously presented for the YZ CMi flare.  

These ULTRACAM observations show that it is possible to use the NBF3510 
and NBF4170 
filters as proxies to understand the components of the white light continuum on timescales not accessible by spectral observations.
Additional observations with ULTRACAM of larger flares (hence, well-measured
flare colors) would provide a more
precise determination of the flare color during all phases of the
flare.  More observations are also critical for finding and characterizing the properties 
of the anti-correlation between continuum components.
 
\acknowledgements We are grateful to Robert Ryans (QUB) for 
installing and maintaining the ULTRACAM pipeline software. AFK 
acknowledges support from NSF grant AST 0807205.
Based on observations obtained with the 4.2-meter William Herschel Telescope operated on the
island of La Palma by the Isaac Newton Group in the Spanish
Observatorio del Roque de los Muchachos of the Instituto de Astrofísica de
 Canarias, and observations with the Apache Point Observatory 3.5-meter telescope, 
which is owned and operated by the Astrophysical Research Consortium.
\bibliography{kowalski_a}

\end{document}